\documentclass[aps,pre,superscriptaddress]{revtex4}
\usepackage{amssymb,amsmath,epsfig}
\begin{document}
\title {Extinction dynamics of Lotka-Volterra ecosystems on evolving networks}
\author{F. Coppex}
\affiliation{Department of Physics, University of Geneva, CH 1211 Geneva 4, Switzerland}
\author{M. Droz}
\affiliation{Department of Physics, University of Geneva, CH 1211 Geneva 4, Switzerland}
\author{A. Lipowski}
\affiliation{Department of Physics, University of Geneva, CH 1211 Geneva 4, Switzerland}
\affiliation{Faculty of Physics, A.~Mickiewicz University, 61-614 Pozna\'{n}, Poland}
 
%%%%%%%%%%%%%%%%%%%%%%%%%%%%%%%%%%%%%%%%%%%%%%%%%%%%%%%%%%%%%%%%%%%%%%%%%%%%%%%
\begin {abstract}
We study a model  of a multi-species ecosystem described by 
Lotka-Volterra-like equations.
Interactions among species form a network whose evolution is determined 
by the dynamics of the model.
Numerical simulations show power-law distribution of intervals between 
extinctions, but only for ecosystems with sufficient variability of 
species and with networks of connectivity above certain threshold that is 
very close to the percolation threshold of the network.
Effect of slow environmental changes on extinction dynamics, degree 
distribution of the network of interspecies interactions, and some 
emergent properties of our model are also examined.
\end{abstract}
\maketitle
%%%%%%%%%%%%%%%%%%%%%%%%%%%%%%%%%%%%%%%%%%%%%%%%%%%%%%%%%%%%%%%%%%%%%%%%%
\section{introduction}
Lotka-Volterra models of interacting species have a  well established 
role in population ecology~\cite{LV}.
Being inspired by an oscillatory behavior  in some prey-predator 
systems, 
they are typically used to model populations on time scale shorter than  
lifetime of describing species.
It means that long-term properties of ecosystems (macro-evolution) are 
usually not captured within such an approach.
On the other hand, models used to describe macro-evolution very often
use the dynamics that operates at the level of species rather than 
individuals.
Such coarse-grained models usually refers to the notion of fitness of a 
species that is not commonly accepted~\cite{BAKSNEPP}.

Recently, there has been some attempts to study macro-evolution using 
models equipped with dynamics that operates at the level of 
individuals~\cite{STAUFFER,ZIA,HALL}.
Taking into account that Lotka-Volterra models are relatively 
successful in describing many aspects of population dynamics it would be 
desirable to apply such an approach also to macroevolution.
Some time ago Abramson introduced a discrete version of Lotka-Volterrra 
ecosystem~\cite{ABRAMSON} and studied certain characteristics of 
extinctions.
His model is an example of a one-dimensional food chain with $M(\sim 
100)$ trophic levels and a single species occupying a given trophic 
level.
Since in realistic food webs $M\sim 4-6$ with typically many species 
belonging to a given trophic level~\cite{MCKANE,QUINCE}, these are highly 
nonrealistic assumptions.
Nevertheless, extinction dynamics in Abramson's model shows some 
features that are characteristic to Earth biosystem.

In the present paper we introduce a Lotka-Volterra model that describes 
a simplified ecosystem of $N$ species of predators and one species of 
preys.
Our model can be thus considered as a simple food web model with only 
two trophic levels.
Competition between predator species is described by a certain 
network~\cite{ALBERT} of interactions whose evolution is coupled with 
dynamics of the model.
Namely, when a certain species becomes extinct (i.e., its density falls 
below a certain threshold) it is replaced by new species with a newly 
created set of interactions with some of existing species.
Despite obvious simplifications the model exhibits some properties that 
are typical to more complicated ecosystems, as for example power-law 
distributions of intervals between extinctions.
Within our model we can also examine how robust this power-law 
distribution is.
We find that under certain conditions, as for example very sparse 
interactions between species, or too strong dominance of a small group of 
species, these power-law characteristics disappear and the model is 
driven into a regime where extinctions have exponential distributions or 
where there are no extinctions and the ecosystem enters a steady state.
In our opinion, such regimes might be relevant when a restricted 
(either in space or time) evolution of an ecosystem or its part is studied.
Interestingly, a threshold value of connectivity that separates 
power-law extinctions and steady state is very close to the percolation 
threshold of the random network of inter-species interactions.

According to a large class of statistical physics
models of biological evolution, avalanches of extinctions do not 
require external factors to trigger them, but might be a natural consequence 
of  the dynamics of an ecosystem.
As a result, these external factors, as e.g., climate changes, solar 
activity or impact of a big meteorite, are very often neglected in such 
studies~\cite{ROBERTS}.
But such factors certainly affect the ecosystem and there is a good 
evidence of it~\cite{NEWMAN}.
Let us emphasize that even the basic mechanism that triggers avalanches 
of extinctions is not known and is a subject of an intensive 
multidisciplinary 
debate~\cite{newman_palmer_book}.

One possibility to take external factor(s) into account in our model is 
to modify a growth rate of prey.
Since dynamics of the model is nonlinear, such a change might have more 
dramatic consequences than merely a change of densities of species.
And indeed we noticed that dynamics of extinctions is strongly 
dependent on the growth rate.
It turns out, that in our model abundance of preys leads to a larger 
frequency of extinctions, and in periods of hunger there are less 
extinctions.
This is clearly due to nonlinearity of the dynamics. 
Larger growth rate increases the density of preys that in turn 
increases densities of predators.
With increased densities, dynamics becomes more competitive and 
extinctions become more frequent.
Such a periodically modulated growth rate leaves some traces also in 
the probability distribution of extinctions.
It might be interesting to notice that paleontological data also show 
some  traces of periodic events, but their proper understanding is still missing~\cite{NEWMAN,RAUP}

During evolution some species are favored and selected at the expense 
of  less fortunate ones.
Evolution constantly searches for best solutions that resembles an 
optimization process.
For example a large size of organisms of a given species might be of 
advantage in some situations, but might cause some problems in the 
other.
What might be Nature's solution to this problem?
Will it be middle-size species or rather two groups of species sitting 
at the extremes of conflicting requirements?
In our opinion, this aspect of evolution is also often omitted in 
models of macroevolution.
Within our model we looked at such emergent properties of species 
selected by evolution.
It turns out that depending on some dynamical details, our model can 
reproduce both types of solutions of such an optimization problem.

In Section II we introduce our model and briefly describe the  
numerical method we used.
Obtained results are presented in Section III.
In Section IV we summarize our results and suggest some further 
extensions of our work.
%%%%%%%%%%%%%%%%%%%%%%%%%%%%%%%%%%%%%%%%%%%%%%%%%%%%%%%%%
\section{Model and numerical calculations}
We study a Lotka-Volterra ecosystem that consists of $N$ species of 
predators with densities $\rho_i\ (i=1,2,\ldots ,N)$ who are all feeding 
on one species of preys with density $\rho_0$.
We assume that each predator species $i$ is characterized by a 
parameter $k_i$ ($0<k_i<1$) that enters evolution equations of the 
model through death and growth terms
\begin{subequations}
\label{lveq}
\begin{eqnarray}
\dot{\rho_0} & = & g(t)\rho_0(1-\rho_0)-\frac{\rho_0}{N}\sum_{i=1}^N 
f(k_i)\rho_i\label{lv1}\\
\dot{\rho_i} & = & 
-d(k_i)\rho_i(1-\rho_0)+f(k_i)\rho_i\rho_0\left(1-\frac{k_i\rho_i+\sum_{j}' k_j\rho_j}{k_i+\sum_{j}' k_j}\right),
\label{lv2}
\end{eqnarray}
\end{subequations}
where $i=1,2,\ldots,N$.
In our model we assume that species interact mainly through 
environmental capacity terms (the last term in Eq.~(\ref{lv2})).
Namely, the growth rate of a given species $i$ is reduced not only due 
to  its density but also due to weighted (with the factor $k$) 
densities of a group of randomly selected neighboring species.
In Eq.~(\ref{lv2}) summation over these neighboring species is denoted 
by ($\sum'$).
Approximately, we might interpret the coefficient $k_i$ as the size of
organisms of $i$-th species -- the bigger they are the bigger their 
role in the environmental capacity term.
We also assume that the growth rate of preys is corrected by the 
environmental capacity term and due to external factors might be a slowly 
varying function of time ($g(t)$).
In principle, external factors might affect also other terms of model 
(\ref{lveq}), but for simplicity we restrict its influence only to the 
growth rate of preys.
Functions $d(k)$ and $f(k)$ reflect the $k$-dependence of death and 
growth of our species.
Explicit form of functions $g(t),\ f(k)$ and $d(k)$ will be given 
later.

Differential equations (\ref{lveq}) are solved using Runge-Kutta 
fourth-order method.
Multi-species Lotka-Volterra ecosystems were subject to intensive 
studies since the pioneering work of May~\cite{MAY}.
It is known that such systems might evolve toward the steady-state with 
positive densities.
However, in some cases, in the steady state density of some species 
might be zero.
Each time a density of a certain species in model (\ref{lveq}) drops 
below a threshold
value which we fix as $\varepsilon=10^{-7}$ we consider such a species 
as extinct~\cite{EPSILON}.
Such a species is then replaced by a new species with a randomly 
assigned density (from the interval (0,1)), the coefficient $k$ ($0<k<1$) 
that is randomly drawn from the distribution 
$p(k)$, and a new set of neighbors (all links of the 'old' species are 
removed).
With such rules the model rather describes $N$ niches, and we assume 
that a time to create a species that will occupy a niche is relatively 
short comparing to the typical lifetime of species.

We assume that a newly created species makes $z$ links with randomly 
selected neighbors.
Links are not directional so a newly created species will also enter 
the evolution equation of species it is neighboring.
If the extinct species would be chosen randomly  the network of 
interactions would have been a random graph.
However, it is the dynamics (\ref{lveq}) that determines which species 
are extinct.
Thus, extinct species are not selected randomly  and the resulting 
network is in general not a random graph.
%%%%%%%%%%%%%%%%%%%%%%%%%%%%%%%%%%%%%%%%%%%%%%%%%%%%%%%%%%
\section{Results}
In the following we describe numerical results obtained for some 
particular cases of model (\ref{lveq}).
\subsection{Intervals between extinctions}
\begin{figure}
\centerline{\epsfxsize=0.7\columnwidth
\epsfbox{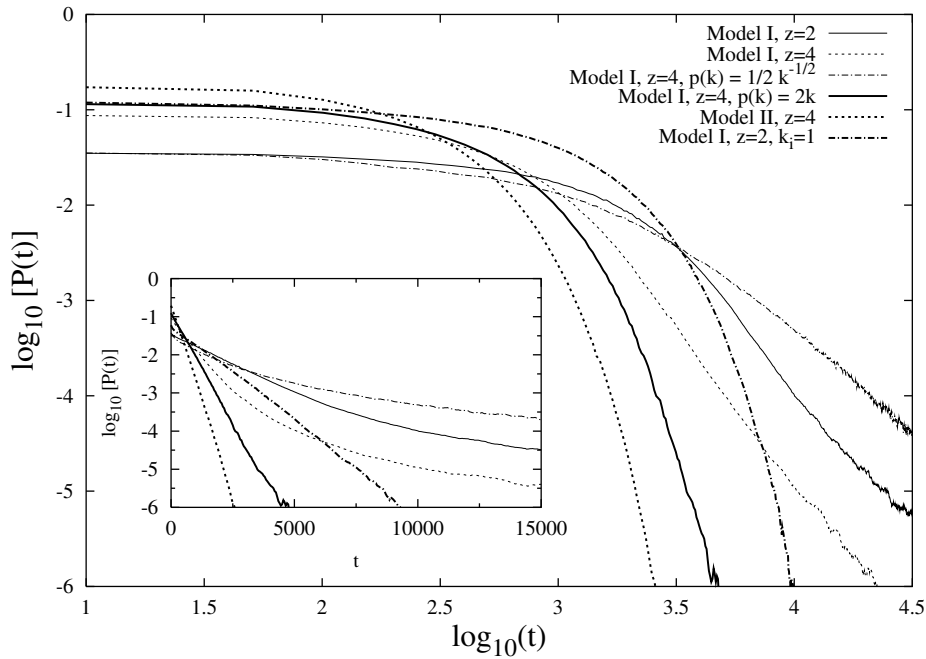}
}
\caption{
Probability distribution of intervals between successive extinctions 
$P(t)$ calculated for some particular cases of model (\ref{lveq}) for 
$N=100$.
Inset shows the same data but plotted on a lin-log scale.
}
\label{intervals}
\end{figure}

Various paleontological data suggest that dynamics of extinctions has 
some power-law distributions of sizes or durations~\cite{NEWMAN}.
In our model we measured time intervals $t$ between successive 
extinctions.
In this calculations we used a constant growth term of preys 
$g(t)\equiv 1$.
We examined two cases: (i) model I: $f(k_i)\equiv 1,\ d(k_i)\equiv 1$ 
and (ii) model II: $f(k_i)=k_i ,\ d(k_i)\equiv 1$.
Unless specified otherwise we select $k_i$ randomly with a homogeneous 
distribution on the interval (0,1) ($p(k)=1$).
Our results are shown in Fig.~\ref{intervals}.
In the simplest case, model I with $z=4$ and $k_i\equiv 1$ (i.e., all 
species during the evolution have identical $k_i(=1)$) we obtain 
exponentially decaying distribution of intervals between extinctions $P(t)$.
Such a decay is also seen for model I (z=4) with linear distribution of $k_i$ namely $p(k)=2k$.
We expect that such a behavior appears when a distribution of $k_i$ in 
the ecosystem is relatively narrow and shifted toward unity.
Such an effect might be due to the small width of distribution $p(k)$ 
(i.e., a distribution from which we draw $k_i$) or might be dynamically 
generated
as in model II.
In this case even though $k_i$ are chosen from a homogeneous 
distribution, the dynamics favors large $k_i$ species (due to their larger growth 
rate) and they dominate the ecosystem.
When the distribution of $k_i$ in the ecosystem is more uniform (model 
I with $p(k)=1$) our simulations suggest that $P(t)$ decays as a power 
law.
Let us notice, however, that a power-law behavior is seen only on 
approximately one decade and we cannot exclude that on a larger time scale a 
different (perhaps exponential) behavior appears as was already 
observed in some other macroevolutionary models~\cite{STAUFFER}.
Let us also notice that for model I with $p(k)=k^{-1/2}/2$ the 
power-law distribution $P(t)$ seems to decay as $t^{-2}$, i.e., with the 
exponent consistent with some paleontological data~\cite{NEWMAN} as well as 
with predictions of some other models~\cite{ZIA}. However, one has to 
recognize that the error bars on experimental data are rather large  and 
that a non-power law behavior cannot be excluded.
%%%%%%%%%%%%%%%%%%%%
%%%%%%%%%%%%%%%%%%%%%%%

\begin{figure}
\centerline{\epsfxsize=0.7\columnwidth
\epsfbox{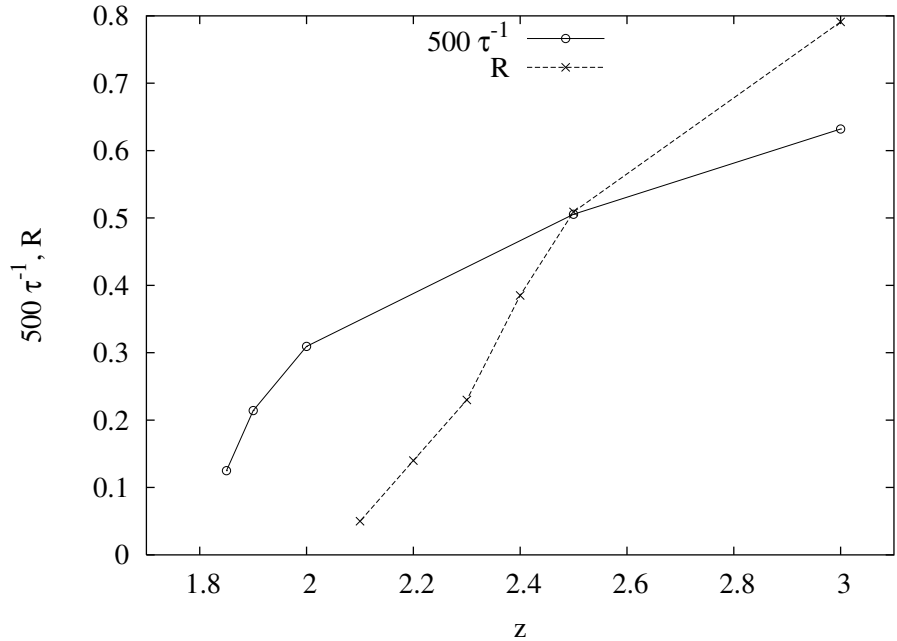}
}
\caption{
The inverse average time between extinctions $\tau^{-1}$ and the 
percolation probability $R$ as  a function of $z$.
Plotted results are based on calculations for $N=100$, 200, 300 and 400 
and extrapolation $N\rightarrow\infty$.
}
\label{tau}
\end{figure}

Note that a power-law decay of $P(t)$ is seen only for sufficiently 
large $z$. When $z$ is too small, we observed that the ecosystem enters the 
steady state where all $\rho_i$ are positive and there are no 
extinctions.
This is probably due to the fact that the competition among predators 
is too weak (or rather too sparse).
To examine the transition between this two regimes in more detail we 
measured the averaged time between extinctions $\tau$ and the results are 
seen in Fig.~\ref{tau}.
One can see that $\tau$ diverges around $z\sim 1.8$~\cite{COMMENT}.
Such a value of the threshold parameter suggests that this transition 
might be related with the percolation transition in our network of 
interspecies interactions.
To examine such a possibility we measured the average size of the 
largest cluster of connected links in the network $R$ (normalized by the 
number of species $N$) and the results are shown in Fig.~\ref{tau}.
Vanishing of this quantity locates the percolation 
transition~\cite{STAUFFERAHARONY}.
One can see that the percolation transition takes place at a larger 
value namely around $z\sim 2.0$.
Our results suggest that these two transitions take place at different 
values of $z$.
However the analysis of finite size effects especially in the 
estimation of $\tau$ is rather difficult and we cannot exclude that these two 
transitions actually overlap, as might be suggested by their proximity.
Such a result would show that the dynamical regime of an ecosystem 
(i.e., steady state or active with power-law distribution of extinctions) 
is determined by the geometrical structure of its interactions.
%%%%%%%%%%%%%%%%%%%%
%%%%%%%%%%%%%%%%%%%%%%%
\subsection{Effect of a modulated growth rate}
\begin{figure}
\centerline{\epsfxsize=0.7\columnwidth
\epsfbox{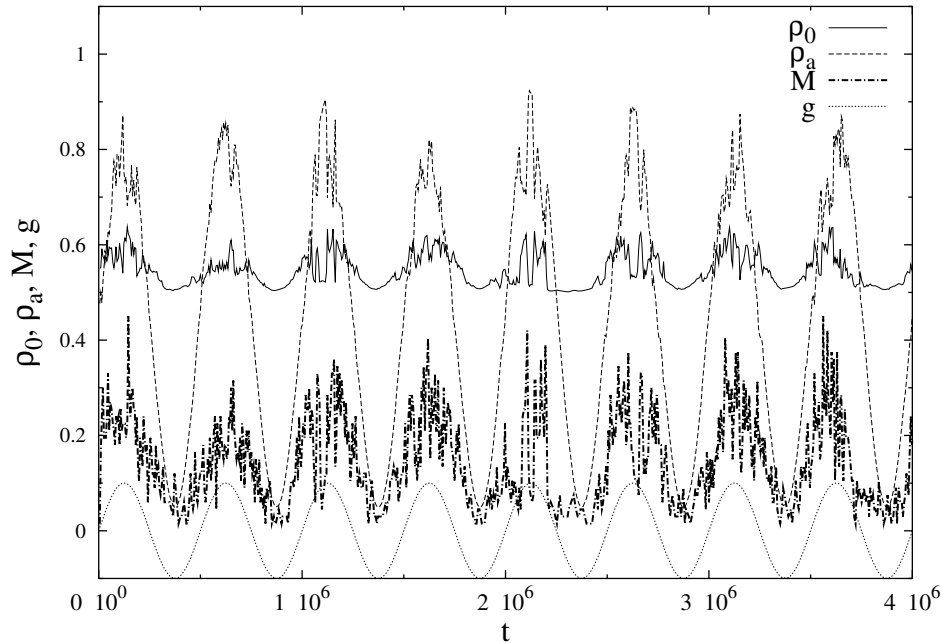}
}
\caption{
A time evolution of the density of preys $\rho_0$,  average density of predators $\rho_a=\frac{1}{N}\sum_{i=1}^{N}$, and the number of extinctions $M$ (divided by 20) in the time interval $\Delta t=10^3$ for the model I with $N=100$ and $z=4$. A rescaled modulated growth rate $(g(t)-1)/10=0.09 \sin(\frac{2\pi t}{T})$ ($T=10^5$) is also shown.
}
\label{modul0}
\end{figure}

Now we examine the role of a modulated in time growth rate of preys.
Such a modulation is supposed to mimic the influence of an external 
factor like a change of a climate.
One of the questions that one can ask in this context is how such a 
change affects the extinction dynamics.
We studied model I with $p(k)=1$ and $d(k_i)\equiv 1$.
The growth rate of preys we chose as $g(t)=1+A{\rm sin}(2\pi t/T)$, 
where 
$A$ and $T$ are parameters.
A typical behavior in case of model I with such a growth rate is shown 
in Fig.~\ref{modul0}.
%%%%%%%%%%%%%%%%%%%%
%%%%%%%%%%%%%%%%%%%%
One can see that increased growth rate increases the density of preys 
$\rho_0$ that increases the density of predators.
However, it increases also the frequency of extinctions.
Such a behavior, namely increased extinction rate during abundance of 
food, might at first sight look as counterintuitive.
This effect is related with the form of environmental capacity terms  
in
in the growth rate in Eq.~(\ref{lv2}), namely
$1-(k_i\rho_i+\sum_{j}' k_j\rho_j)/(k_i+\sum_{j}' k_j)$.
Such term certainly has a larger variability for increased density of 
predators $\rho_i$, and for some species (depending on the distribution 
of links, coefficients $k_i$ and densities) it causes faster 
extinction.
Let us also notice that since period of modulation $T$ is quite large, 
there is no retardation effect between density of preys and predators.
We observed such retardation for smaller values of $T(\sim 1000)$.

\begin{figure}
\centerline{\epsfxsize=0.7\columnwidth
\epsfbox{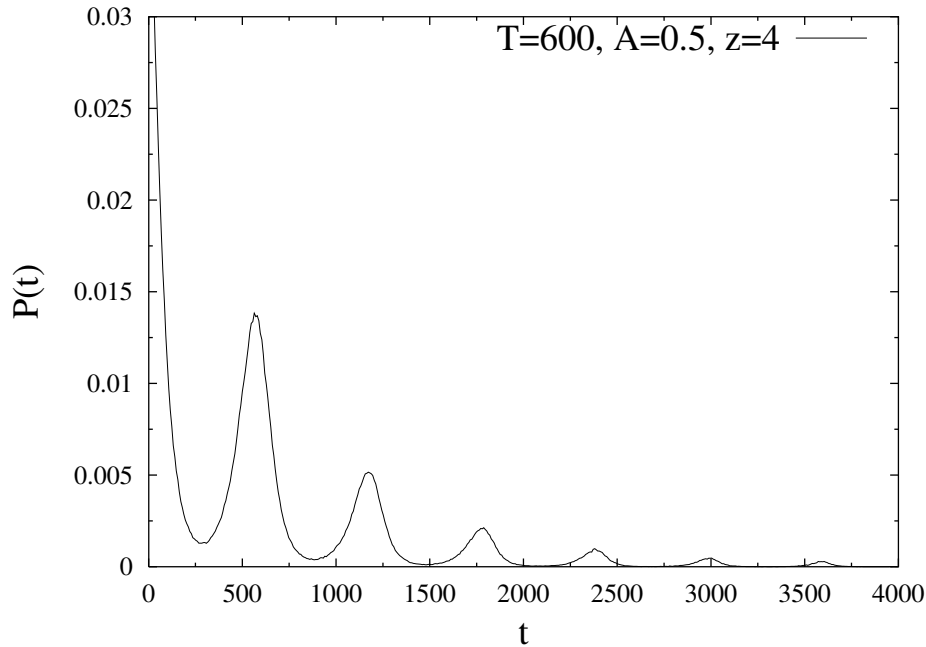}
}
\caption{
Probability distribution of intervals between successive extinctions 
$P(t)$ calculated for model I with modulated growth rate ($N=100$).
}
\label{modul}
\end{figure}

Modulated growth rate of prays affects also the probability 
distribution of intervals between extinctions $P(t)$ as shown in 
Fig.~\ref{modul}.
One can see that period of modulation $T$ is imprinted in $P(t)$.
Let us notice that certain paleontological data do show some signs of 
periodicity but its origin still remains unclear~\cite{RAUP,NEWMAN}.

It is known that slowly changing ecosystems sometimes undergo 
catastrophic shifts~\cite{SCHEFFER}.
As a result, the ecosystem switches to a contrasting alternative stable 
state.
It would be interesting to examine whether multi-species ecosystems, as 
described by our model (\ref{lveq}), might also exist in such 
alternative states.
If so, one can ask whether for example structure of the network of 
interspecies interactions or extinction dynamics are the same in such 
states.
%%%%%%%%%%%%%%%%%%%%
%%%%%%%%%%%%%%%%%%%%%%%
\subsection{Emergent properties of species}
\begin{figure}
\centerline{\epsfxsize=0.7\columnwidth
\epsfbox{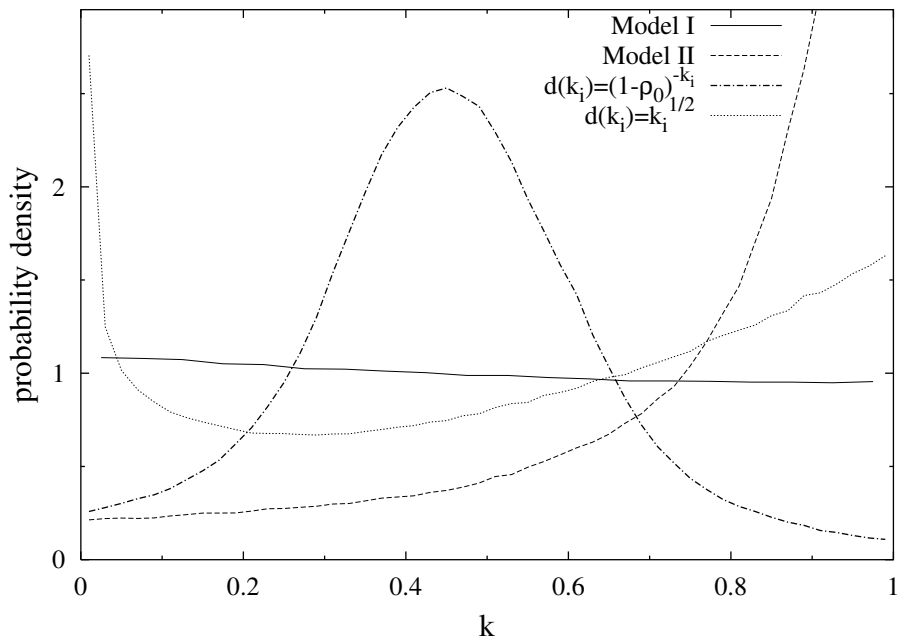}
}
\caption{
Distribution of $k_i$ in the steady state of some particular cases of 
model
(\ref{lveq}) (see text).
}
\label{emer}
\end{figure}

It might be interesting to ask what are the characteristics of species 
that are preferred by the evolution in our ecosystem.
Since species are characterized only by the number $k_i$ it is 
equivalent to calculating the distribution  of $k_i$ in the steady state of 
model (\ref{lveq}).
Of course, due to selection this distribution in general will be 
different than the distribution $p(k)$, i.e., the distribution from which we 
draw 
$k_i$ of a newly created species.
Some of our results are shown in Fig.~\ref{emer} (all results for 
$g(t)\equiv 1$,\ z=4,\ N=100).

In the case of model I ($f(k_i)\equiv 1,\ d(k_i)\equiv 1$) with 
homogeneous initial distribution of $k_i$ ($p(k)=1$) one can see that the 
steady state distribution is also approximately homogeneous
(with a slight bias favoring small-$k$ species).
We checked that model I shows this behavior also for other 
distributions $p(k)$ (what you put is what you get).
Different behavior appears for model II ($f(k_i)=k_i,\ d(k_i)\equiv 
1$).
In this case the growth rate factor $f(k_i)$ of $i$-th species is 
proportional to $k_i$ that certainly prefers species with large $k_i$.
Numerical results for homogeneous distribution $p(k)=1$ confirm such a 
behavior (Fig.~\ref{emer}).
We observed similar strong preference of large $k_i$ species also for 
Model II with other distributions $p(k)$.

We also examined the selection pattern in presence of some competing 
effects.
To compensate a strong preference toward large-$k$ species we made 
simulations for our model with $f(k_i)=k_i,\ d(k_i)=\sqrt{k_i}$ and 
$p(k)=1$.
Such term reduces the death rate of small-$k$ species.
Our results show (Fig.~\ref{emer}) that in this case distribution of 
$k_i$ has two maxima at the extremities of the interval (0,1).
On the other hand with the same model but for 
$d(k_i)=(1-\rho_0)^{-k_i}$ (that also reduces the death rate of small-$k$ species) we obtain a 
distribution with a single maximum around $k=0.45$.
It would be desirable to understand the origin of the qualitative 
difference between these two cases.
%%%%%%%%%%%%%%%%%%%%
%%%%%%%%%%%%%%%%%%%%%%%

Actually, there is yet another property of our species that is 
subjected to evolutionary selection, namely the number of links $l_i$ (degree) 
of a given species.
Although at the beginning each species acquires $z$ links this number 
changes during the evolution because some links of a given species 
might be created or removed due to creation or extinction of another 
species.
And since it is the dynamics of our model and not the random process 
that determines which species are removed, one can expect that the degree 
distribution might be different from the Poissonian distribution that 
is characteristic for random graphs (see~\cite{BOLLO} for a precise 
definition of random graphs).

\begin{figure}
\centerline{\epsfxsize=0.7\columnwidth
\epsfbox{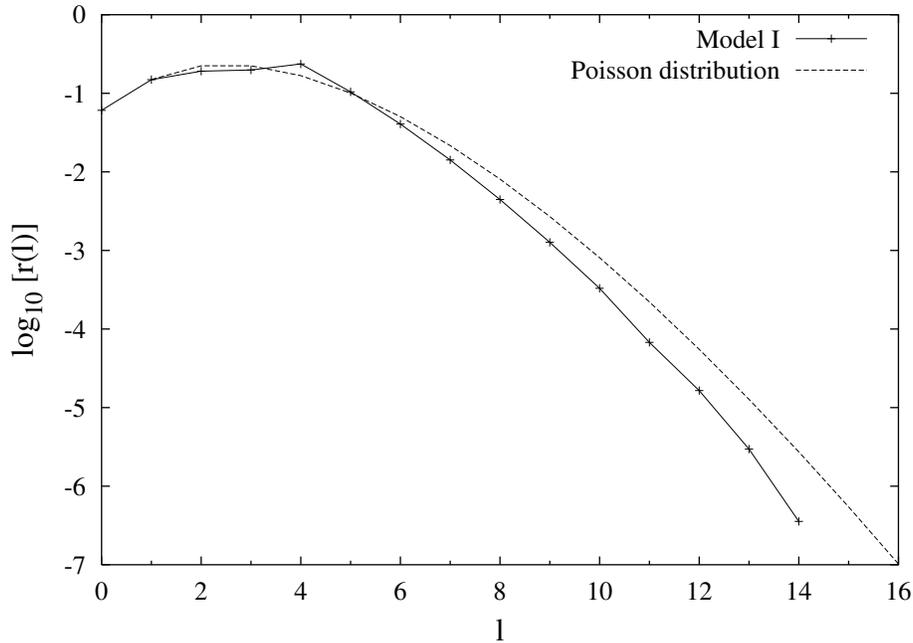}
}
\caption{
Probability distribution $r(l)$ of sites with a given connectivity $l$ 
for model I with $z=4$ and $N=100$ compared with the corresponding 
Poissonian distribution ($\langle l_i \rangle = 2.98$).
}
\label{link}
\end{figure}

To check statistical properties of the network of interactions in our 
model we calculated the degree distribution.
Our results for model I with $z=4$ and $N=100$ are shown in 
Fig.~\ref{link}.
Let us notice that although each species has $z$ links at the 
beginning
it does not mean that the average number of links connected to a given 
site $\langle l_i \rangle$ equals $z$ since dynamics of the model might 
preferentially remove sites of certain connectivity.
And indeed, numerical calculations show that in this case $\langle 
l_i\rangle =2.98<z=4$, i.e., dynamics preferentially removes sites of 
larger connectivity.
For comparison with the random graph we plot also the Poissonian 
distribution
$r(l)={\rm e}^{-\langle l_i \rangle}\langle l_i \rangle^l/l!$, where 
$\langle l_i \rangle=2.98$. It should be emphasized that the distribution 
might be approximately fitted  using a Poisson distribution, for 
example with $\langle l_i \rangle=2.65$. However, it is then not a physically 
relevant distribution since the average connectivity $\langle l_i 
\rangle=2.65$ differs from the value $\langle l_i \rangle=2.98$ obtained 
from the simulations. In this sense the distribution is not Poissonian.
One can see that for large connectivity the degree distribution decays 
faster than the Poissonian distribution.
This result confirms that dynamics of the model preferentially removes 
highly connected species.
Such sites are probably more susceptible to fluctuations in the system 
due to extinctions and creation of new species.
On the other hand, poorly connected species are more likely to arrive 
at a relatively stable state.
Similar results concerning the degree distribution were obtained in 
some other cases of our model.
%%%%%%%%%%%%%%%%%%%%
%%%%%%%%%%%%%%%%%%%%%%%
%%%%%%%%%%%%%%%%%%%%%%%%%%%%%%%%%%%%%%%%%%%%%%%%%%%%%%%%%%%%%%%%%%%%
\section{Conclusions}
In the present paper we studied extinction dynamics of a Lotka-Volterra 
model of a two-level food web.
In our model $N$ species of predators feed on a single species of 
preys.
Competition between predators, that is specified by a certain network 
of interactions,  leads to their extinction and replacement by new 
species.
Distribution of intervals between successive extinctions in some cases 
has power-law tails and thus resembles extinction pattern of the real 
ecosystem.
However, when the network of interactions between predators is too 
sparse the ecosystem enters the steady state.
We have shown that such a change of behavior might be related with a 
percolation transition of the network.
We also examined an influence of external factors on the evolution of 
the ecosystem.
More specifically, we studied the evolution of our model in case when 
the growth rate of preys is changing periodically in time.
It turns our that such a modulation substantially changes the frequency 
of extinctions.
Counterintuitively, periods with abundance of preys have higher 
frequency of extinctions than periods with lesser amount of preys.
Moreover, we examined some properties of species that are 
preferentially selected by the dynamics of our model.
Under some conditions preferred species are a compromise to the 
conflicting dynamics.
Under some other conditions, preferred species form two antagonistic 
(with respect to the conflicting rules) groups.
We also examined the degree distribution of the network of interactions 
between species.
It turns out that dynamics of the model has a slight preference to 
remove species of higher connectivity.
As a result degree distribution shows some deviation from Poissonian 
distribution that is characteristic to random graphs.

It would be desirable to examine some extensions of our model.
For example one can introduce additional trophic levels or other forms 
of interspecies interactions.
One can also examine a variable number of species that would allow to 
create new species using certain mutation mechanism rather than assuming 
that they appear as soon as a niche becomes empty.
Another possibility that is outside the scope of majority of 
macroevolutionary models would be to make further study of emergent properties of 
species.
For example, one can imagine that a group of species in the ecosystem 
is well adapted and essentially not subjected to evolutionary changes.
On the other hand there is a group of 'newcomers' where evolutionary 
changes are much more frequent.
How evolution and properties of 'newcomers' are influenced by the 
properties of well-adapted species?
Such problems might be easily approached within our model.
Selection of a certain group of species (with a given value of $k$ for 
example) can be considered as a selection of a certain strategy.
One can examine models of this kind where species are have 
multi-component parameters [$k=(k^a,k^b,\ldots)$].
Consequently, one can study evolutionary selection of more complicated 
traits, strategies, or behaviors.
Such an approach would provide an interesting link with certain 
evolutionary aspects of game theory~\cite{GAME}.

This work was partially supported by the Swiss National Science 
Foundation
and the project OFES 00-0578 "COSYC OF SENS".
Some of our calculations were done on 'openMosix Cluster' built and
administrated by Lech D\c{e}bski at the Institute of Physics at the 
Adam Mickiewicz University (Poland).
%%%%%%%%%%%%%%%%%%%%%%%%%%%%%%%%%%%%%%%%%%%%%%%%%%%%%%%%%%%%%%%%%%%%%%%%%%%%%%

%%%%%%%%%%%%%%%%%%%%%%%%%%%%%%%%%%%%%%%%%%%%%%%%%%%%%%%%%%%%%%%%%%%%%%%%%%%%%%%
%%%%%%%%%%%%%%%%%%%%%%%%%%%%%%%%%%%%%%%%%%%%%%%%%%%%%%%%%%%%%%%%%%%%%%%%%%%%%%%

\begin{thebibliography}{}
\bibitem{EPSILON} Statistical properties of extinctions that we study 
in this paper are not affected by the precise value of the $\varepsilon$ 
as soon as it has a small value ($\varepsilon\ll 1$).
\bibitem{COMMENT} For noninteger $z$ (i.e., the number of links of 
newly created species) we used the following procedure: with probability 
$z-[z]$ we created $[z]+1$ links and with probability $1+[z]-z$ we 
created 
$[z]$ links.
On average such a recipe produces $z$ links.
\end{thebibliography}
\end {document}